# Hyperbolic metamaterial interfaces: Hawking radiation from Rindler horizons and the "end of time"

Igor I. Smolyaninov [1)], Ehren Hwang [1)], and Evgenii Narimanov [2)]

[1)] *Department of Electrical and Computer Engineering, University of Maryland, College Park, MD 20742, USA*

[2)] *Birck Nanotechnology Centre, School of Electrical and Computer Engineering, Purdue University, West Lafayette, IN 47907, USA*

**Extraordinary rays in a hyperbolic metamaterial behave as particle world lines in a three dimensional (2+1) Minkowski spacetime. We analyze electromagnetic field behavior at the boundaries of this effective spacetime depending on the boundary orientation. If the boundary is perpendicular to the space-like direction in the metamaterial, an effective Rindler horizon may be observed which produces Hawking radiation. On the other hand, if the boundary is perpendicular to the time-like direction an unusual physics situation is created, which can be called "the end of time". It appears that in the lossless approximation electromagnetic field diverges at the interface in both situations. Experimental observations of the "end of time" using plasmonic metamaterials confirm this conclusion.**

Modern developments in gravitation research strongly indicate that classic general relativity is an effective macroscopic field theory, which needs to be replaced with a more fundamental theory based on yet unknown microscopic degrees of freedom. On the other hand, our ability to obtain experimental insights into the future fundamental theory is strongly limited by low energy scales available to terrestrial particle physics and astronomical observations. The emergent analogue spacetime program offers a promising way around this difficulty. Looking at such systems as superfluid helium and cold atomic Bose-Einstein condensates, physicists learn from Nature and discover how macroscopic field theories arise from known well-studied atomic degrees of freedom. An interesting recent example of this approach is Horava gravity [1], which is based on the well known Lifshitz point behavior in solid state physics. Another exciting development along this direction is recent introduction of metamaterials and transformation optics [2-4]. The latter field is not limited by the properties of atoms and molecules given to us by Nature. “Artificial atoms” used as building blocks in metamaterial design offer much more freedom in constructing analogues of various exotic spacetime metrics, such as black holes [5-9], wormholes [10,11], spinning cosmic strings [12], and even the metric of Big Bang itself [13]. Explosive development of this field promises new insights into the fabric of spacetime, which cannot be gleaned from any other terrestrial experiments.

On the other hand, compared to standard general relativity, metamaterial optics gives more freedom to design an effective space-time with very unusual properties. Light propagation in all static general relativity situations can be mimicked with positive $\varepsilon_{ik} = \mu_{ik}$ [14], while the allowed parameter space of the metamaterial optics is broader. Thus, flat Minkowski space-time with the usual (-,+,+,+) signature does not need to be a starting point. Other effective signatures, such as the “two times” (2T) physics (-,-,+,+) signature may be realized [15]. Theoretical investigation of the 2T higher dimensional space-time models had been pioneered by Paul Dirac [16]. More recent examples can be found in [17,18]. Metric signature change events (in which a phase transition occurs between say (-,+,+,+) and (-,-,+,+) space-time signature) are being studied in Bose-Einstein condensates and in some modified gravitation theories (see ref.[19], and the references therein). It is predicted that a quantum field theory residing on a spacetime undergoing a signature change reacts violently to the imposition of the signature change. Both the total number and the total energy of the particles generated in a signature change event are formally infinite [19]. While optics of bulk hyperbolic metamaterials provides us with ample opportunities to observe metric signature transitions [15], even more interesting physics arise at the metamaterial interfaces. Very recently it was demonstrated that mapping of monochromatic extraordinary light distribution in a hyperbolic metamaterial along some spatial direction may model the "flow of time" in a three dimensional (2+1) effective Minkowski spacetime [13]. If an interface between two metamaterials is engineered so that the effective metric changes signature across the interface, two possibilities may arise. If the interface is perpendicular to the time-like direction $z$, this coordinate does not behave as a "timelike" variable any more, and the continuous "flow of time" is interrupted. This situation (which cannot be realized in classic general relativity) may be called the "end of time". It appears that optics of metamaterials near the "end of time" event is quite interesting and deserves a detailed study. For example, in the lossless approximation all the possible "end of time" scenarios lead to field divergencies, which indicate quite interesting linear and nonlinear optics behaviour near the "end of time". On the other hand, if the metamaterial interface is perpendicular to the space-like direction of the effective (2+1) Minkowski spacetime, a Rindler horizon may be observed (Rindler metric approximates spacetime behaviour near the black hole event horizon [14]).

Let us begin with a brief summary of refs.[13,15], which demonstrated that a spatial coordinate may become "timelike" in a hyperbolic metamaterial. To better understand this effect, let us start with a non-magnetic uniaxial anisotropic material with dielectric permittivities $\varepsilon_x = \varepsilon_y = \varepsilon_1$ and $\varepsilon_z = \varepsilon_2$, and assume that this behaviour holds in some frequency range around $\omega=\omega_0$. Any electromagnetic field propagating in this material can be expressed as a sum of the "ordinary" and "extraordinary" contributions, each of these being a sum of an arbitrary number of plane waves polarized in the "ordinary" ($\vec{E}$ perpendicular to the optical axis) and "extraordinary" ($\vec{E}$ parallel to the plane defined by the k–vector of the wave and the optical axis) directions. Let us define our "scalar" extraordinary wave function as $\varphi=E_z$ so that the ordinary portion of the electromagnetic field does not contribute to $\varphi$. Since metamaterials exhibit high dispersion, let us work in the frequency domain and write the macroscopic Maxwell equations as [20]

$$\frac{\omega^2}{c^2}\vec{D}_\omega = \vec{\nabla}\times\vec{\nabla}\times\vec{E}_\omega \text{ and } \vec{D}_\omega = \ddot{\varepsilon}_\omega \vec{E}_\omega \tag{1}$$

Eq.(1) results in the following wave equation for $\varphi_\omega$ if $\varepsilon_1$ and $\varepsilon_2$ are kept constant inside the metamaterial:

$$-\frac{\omega^2}{c^2}\varphi_\omega = \frac{\partial^2\varphi_\omega}{\varepsilon_1 \partial z^2} + \frac{1}{\varepsilon_2}\left(\frac{\partial^2\varphi_\omega}{\partial x^2} + \frac{\partial^2\varphi_\omega}{\partial y^2}\right) \tag{2}$$

While in ordinary crystalline anisotropic media both $\varepsilon_1$ and $\varepsilon_2$ are positive, this is not necessarily the case in metamaterials. In hyperbolic metamaterials [21] $\varepsilon_1$ and $\varepsilon_2$ have opposite signs. These metamaterials are typically composed of multilayer metal-dielectric or metal wire array structures, as shown in Fig.1. Optical properties of such metamaterials are quite extraordinary. For example, there is no usual diffraction limit in a hyperbolic metamaterial [22,23].

Let us consider the case of constant $\varepsilon_1$ >0 and $\varepsilon_2$ <0, and assume that this behavior holds in some frequency range around $\omega=\omega_0$. Let us assume that the metamaterial is illuminated by coherent CW laser field at frequency $\omega_0$, and we study spatial distribution of the extraordinary field $\varphi_\omega$ at this frequency. Under these assumptions equation (2) can be re-written in the form of 3D Klein-Gordon equation describing a massive scalar $\varphi_\omega$ field:

$$-\frac{\partial^2\varphi_\omega}{\varepsilon_1\partial z^2}+\frac{1}{|\varepsilon_2|}\left(\frac{\partial^2\varphi_\omega}{\partial x^2}+\frac{\partial^2\varphi_\omega}{\partial y^2}\right)=\frac{{\omega_0}^2}{c^2}\varphi_\omega=\frac{m^{*2}c^2}{\hbar^2}\varphi_\omega \tag{3}$$

in which the spatial coordinate $z=\tau$ behaves as a “timelike” variable. Therefore, eq.(3) describes world lines of massive particles which propagate in a flat (2+1) Minkowski spacetime. When a metamaterial is built and illuminated with a coherent extraordinary CW laser beam, the stationary pattern of light propagation inside the metamaterial represents a complete “history” of a toy (2+1) dimensional spacetime populated with particles of mass $m^*$. This “history” is written as a collection of particle world lines along the “timelike” $z$ coordinate. Note that in the opposite situation in which $\varepsilon_1$ <0 and $\varepsilon_2$ >0, equation (2) would describe world lines of tachyons [24] having “imaginary” mass $m^*=i\mu$.

The world lines of particles described by eq.(3) are straight lines, which is easy to observe in the experiment [13]. If adiabatic variations of $\varepsilon_1$ and $\varepsilon_2$ are allowed inside the metamaterial, world lines of massive particles in some well known curvilinear spacetimes can be emulated, including the world line behavior near the "beginning of time" at the moment of Big Bang [13]. Thus, mapping of monochromatic extraordinary light distribution in a hyperbolic metamaterial along some spatial direction may model the "flow of time" in an effective three dimensional (2+1) spacetime. Since the parameter space of metamaterial optics is broader than the parameter space of general relativity, we can also engineer the "end of time" event if an interface between two metamaterials is prepared so that the effective metric changes signature at the interface. In such a case the spatial coordinate does not behave as a "timelike" variable any more, and the continuous "flow of time" is suddenly interrupted, as shown in Fig.2(a). This situation (which cannot be realized in classic general relativity) may be called the "end of time". It appears that optics of metamaterials near the "end of time" event is quite interesting and deserves a detailed study.

For the sake of simplicity let us consider the case of constant $\varepsilon_1=\varepsilon_x=\varepsilon_y>0$ and assume that finite $\varepsilon_2=\varepsilon_z$ changes sign from $\varepsilon_2<0$ to $\varepsilon_2>0$ as a function of $z$ in some frequency range around $\omega=\omega_0$ (see Fig.2(a)). Taking into account $z$ derivatives of $\varepsilon_2$, eq.(1) results in the following equation for $\varphi_\omega$:

$$-\frac{\partial^2\varphi_\omega}{\varepsilon_1\partial z^2}+\frac{1}{(-\varepsilon_2)}\left(\frac{\partial^2\varphi_\omega}{\partial x^2}+\frac{\partial^2\varphi_\omega}{\partial y^2}\right)-\frac{2}{\varepsilon_1\varepsilon_2}\left(\frac{\partial\varepsilon_2}{\partial z}\right)\left(\frac{\partial\varphi_\omega}{\partial z}\right)-\frac{\varphi_\omega}{\varepsilon_1\varepsilon_2}\left(\frac{\partial^2\varepsilon_2}{\partial z^2}\right)=\frac{\omega_0^{\,2}}{c^2}\varphi_\omega \tag{4}$$

Let us consider a plane wave solution in the *xy* direction with an in-plane wave vector *k*. Now we may introduce a new "wave function" $\psi$ as $\varphi_\omega=-\psi/\varepsilon_2$ and obtain

$$-\frac{\partial^2\psi}{\partial z^2}+\frac{\varepsilon_1 k^2}{\varepsilon_2}\psi=\frac{\varepsilon_1\omega_0^{\,2}}{c^2}\psi\ , \tag{5}$$

where the second term may be considered as an effective "potential energy" (note that $\psi$ is proportional to the *B* field of the plane wave). In the region where both $\varepsilon_1$ and $\varepsilon_2$ are positive, the material behaves as a normal anisotropic dielectric. The "end of time" event (Fig.2(a)) in a lossless metamaterial with finite dielectric permittivity may only occur if $\varepsilon_2$ changes sign continuously via $\varepsilon_2$=0. Examples of such transitions may be found in [25]. Let us consider the case of $\varepsilon_2$ changing sign via $\varepsilon_2$ =0 at $z=0$. Unless there is a special physics reason, linear behaviour $\varepsilon_2=\gamma z$ may be assumed near zero. Neglecting the finite right hand side, and after rescaling eq.(5) may be re-written as

$$-\frac{\partial^2 g}{\partial \xi^2}+\frac{1}{\xi}g(\xi)=0 \ , \tag{6}$$

The general solution of eq.(6) is the Bessel-Clifford function: $g\sim C_{-1}(\xi)=\xi^{1/2}I_{-1}(2\xi^{1/2})$ [26]. Near zero $g$~const, and therefore $\varphi_\omega\sim g/\varepsilon_2\sim 1/z$ diverges near $z$=0. This field divergence is somewhat similar to the field divergence predicted for the positive-negative index metamaterial interface considered in [27]. Our numerical simulations of this situation using COMSOL Multiphysics 3.4 solver are presented in Fig.3. These results are consistent with the analytical analysis of the problem. Thus, it appears that in the lossless approximation field divergence is unavoidable in all the possible "end of time" scenarios. This field divergence leads to divergence of nonlinear optical effects at the interface, including higher harmonic generation. In the language of particle physics higher harmonic generation must be interpreted as particle creation at the "moment" $z=\tau=0$ of metric signature change. Thus, our results are also consistent with the analysis of signature change events in ref.[19].

Now let us consider the case of a hyperbolic metamaterial interface, which is oriented perpendicular to the "space-like" direction (Fig.2(b)). For the sake of simplicity, let us consider the case of constant $\varepsilon_2=\varepsilon_z$ <0 and assume that finite $\varepsilon_1(x)=\varepsilon_x=\varepsilon_y$ changes sign from $\varepsilon_1$ >0 to $\varepsilon_1$ <0 as a function of $x$ in some frequency range around $\omega=\omega_0$. Because of translational symmetry along the $z$ direction, we may consider a plane wave solution in the $z$ direction with a wave vector component $k_z$. Introducing $\psi=B$ as above, we obtain

$$-\frac{\partial^2\psi}{\partial x^2}+\frac{\varepsilon_2 k_z^{\ 2}}{\varepsilon_1}\psi=\frac{\varepsilon_2\omega_0^{\ 2}}{c^2}\psi \ , \tag{7}$$

The same analysis as above indicates that $E_x \sim -\frac{1}{\varepsilon_x}\frac{\partial\psi}{\partial z}$ diverges at the interface. Note that the choice of $\varepsilon_1=\alpha x^2$ (where $\alpha>0$) at $x$=0 leads to Rindler-like optical space near the interface. The Rindler metric can be written as

$$ds^2=-\frac{g^2x^2}{c^2}dt^2+dl^2 \quad , \tag{8}$$

where the horizon is located at $x$=0 [28]. The spatial line element of the corresponding Fermat metric as perceived by the Rindler observers is

$$dl^2=\frac{dl^2}{-g_{00}}=\frac{c^4dl^2}{g^2x^2} \tag{9}$$

Since the extraordinary photon wave vector $k\approx k_x \sim \left(-\varepsilon_2/\varepsilon_1\right)^{1/2}k_z$ diverges at the interface (see eq.(7)), the "optical length" element experienced by the extraordinary photons also diverges:

$$dl_{opt}^{2} = \frac{k^2 c^2 dl^2}{\omega_0{}^2} \sim \frac{dl^2}{x^2} \quad , \tag{10}$$

where *dl* is the length element in the coordinate space. Comparison of eqs.(9) and (10) demonstrates that a region of "optical space" near $x=0$ does look like an electromagnetic black hole and $\alpha \sim g^2$ defines effective surface gravity at the horizon. However, we should emphasize that metamaterial losses lead to appearance of the imaginary part in the effective potential $V = \frac{\varepsilon_2 k_z{}^2}{\varepsilon_1}$ in eq.(7), so that no true horizon appears. An attempt to compensate losses with gain, and achieve the "true horizon" would lead to effective Hawking radiation [28] from the interface. Due to divergent density of states in the hyperbolic band of the metamaterial [29], energy pumped into the metamaterial from the outside space would mostly go into the hyperbolic band around $\omega_0$. In the language of effective (2+1) Minkowski spacetime described by eq.(3) new "particles" would be created at the boundary of this spacetime in the layer $\Delta x \sim 1/\alpha^{1/2}$. Due to uncertainty principle, the momentum uncertainty of these particles is $\Delta p_x \sim \hbar \alpha^{1/2}$. At given $\omega_0$ this uncertainty translates into the same uncertainty $\Delta p_z \sim \hbar \alpha^{1/2} \sim T$ of the *z* component of the photon momentum, which plays the role of effective energy in the (2+1) Minkowski spacetime (see Fig.1(c)). Thus, we recover the well known behaviour of the Unruh-Hawking effect:

$$\Delta p_z \sim T_H \sim \hbar \alpha^{1/2} \sim \hbar g \tag{11}$$

Let us consider a possible experimental realization of these effects in the layered metamaterial structure presented in Fig.1(b). Let us assume that the metallic layers are oriented perpendicular to *z* direction. The diagonal components of the permittivity tensor in this case have been calculated in ref. [25] using Maxwell-Garnett approximation:

$$\varepsilon_1 = \alpha\varepsilon_m + (1-\alpha)\varepsilon_d \quad , \qquad \varepsilon_2 = \frac{\varepsilon_m \varepsilon_d}{(1-\alpha)\varepsilon_m + \alpha\varepsilon_d} \tag{12}$$

where $\alpha$ is the fraction of metallic phase, and $\varepsilon_m<0$ and $\varepsilon_d>0$ are the dielectric permittivities of metal and dielectric, respectively. We would like to arrange an “end of time” interface as a function of $\alpha(z)$. As described above, we would like to keep $\varepsilon_1$ positive, while changing sign of $\varepsilon_2$. Simple analysis of eqs.(12) indicates that the “end of time” occurs around $\alpha_0=\varepsilon_m/(\varepsilon_m-\varepsilon_d)$, while we need to keep $\varepsilon_d>-\varepsilon_m$ at the experimental frequency $\omega_0$. Thus, by gradually increasing $\alpha$ through the $\alpha_0$ value we will observe the effect in question. This can be achieved by gradual increase of the metal layer thickness $d_1$, while keeping the dielectric layer thickness $d_2$ constant, so that the necessary range of $\alpha=d_1/(d_1+d_2)$ is achieved. The Rindler horizon can be achieved in a similar manner.

While realization of experimental geometry described above would require complicated nanofabrication, a simpler experiment can be performed using PMMA-based plasmonic hyperbolic metamaterials described in detail in ref. [22]. Rigorous theoretical description of these metamaterials has been developed in ref. [30]. However, the following qualitative model may guide us in understanding similarity between the layered 3D hyperbolic metamaterials shown in Fig.1(b) and plasmonic hyperbolic metamaterials (Fig.4), which are based on PMMA stripes deposited onto gold film surface. Let us consider a surface plasmon (SP) wave which propagates over a flat metal-dielectric interface. If the metal film is thick, the SP wave vector is defined by expression

$$k_p = \frac{\omega}{c}\left(\frac{\varepsilon_d \varepsilon_m}{\varepsilon_d + \varepsilon_m}\right)^{1/2} \tag{13}$$

where $\varepsilon_m(\omega)$ and $\varepsilon_d(\omega)$ are the frequency-dependent dielectric constants of the metal and dielectric, respectively [31]. Let us introduce an effective 2D dielectric constant $\varepsilon_{2D}$, which characterizes the way in which SPs perceive the dielectric material deposited onto the metal surface. Similar to the 3D case, we can introduce $\varepsilon_{2D}$ so that $k_p=\varepsilon_{2D}^{1/2}\omega/c$, and thus

$$\varepsilon_{2D} = \left(\frac{\varepsilon_d \varepsilon_m}{\varepsilon_d + \varepsilon_m}\right) \tag{14}$$

Now it is easy to see that depending on the frequency, SPs perceive the dielectric material bounding the metal surface in drastically different ways. At low frequencies $\varepsilon_{2D} \approx \varepsilon_d$. Therefore, plasmons perceive a PMMA stripe as dielectric. On the other hand, at high enough frequencies around $\lambda_0$~500 nm, $\varepsilon_{2D}$ changes sign and becomes negative since $\varepsilon_d(\omega) > -\varepsilon_m(\omega)$. As a result, around $\lambda_0$~500 nm plasmons perceive PMMA stripes on gold as if they are "metallic layers", while gold/vacuum portions of the interface are perceived as "dielectric layers". Thus, at these frequencies plasmons perceive a PMMA stripe pattern from Fig.4 as a layered hyperbolic metamaterial shown in Fig.1(b).

Rigorous description of plasmonic hyperbolic metamaterials in terms of Diakonov surface plasmons [30] produces similar results. In this description a PMMA grating on the gold film surface is treated as an anisotropic dielectric medium having the following perpendicular and parallel components (defined with respect to the optical axis.) of the diagonal dielectric tensor:

$$\varepsilon_{\perp} = \alpha\varepsilon_d + (1-\alpha) \quad , \quad \varepsilon_{II} = \frac{\varepsilon_d}{(1-\alpha)\varepsilon_d + \alpha} \tag{15}$$

where $\varepsilon_d$ is the permittivity of PMMA, and $\alpha=d_1/(d_1+d_2)$ is defined by the widths $d_1$ and $d_2$ of the PMMA and vacuum stripes, respectively. In the frequency range below $\varepsilon_m(\omega)=-\varepsilon_{\perp}$ plasmon dispersion has normal elliptic character. On the other hand, in the frequency range between $\varepsilon_m(\omega)=-\varepsilon_{\perp}$ and $\varepsilon_m(\omega)=-(\varepsilon_{\perp}\varepsilon_{II})^{1/2}$ the plasmon dispersion is hyperbolic. At $\alpha$ = 0.5 this hyperbolic band is located between $\varepsilon_m(\omega)$ = - 1.63 and $\varepsilon_m(\omega)$= - 1.5. According to material parameters of gold reported in [32], this frequency range is located around $\lambda$=490 nm. Thus, the "end of time" conditions can be achieved in this frequency range via continuous variation of $d_1$ through the $\alpha$=0.5 value in a plasmonic hyperbolic metamaterial.

Fabrication of such plasmonic hyperbolic metamaterials in two dimensions requires only very simple and common lithographic techniques. As described in the experimental scenario above, the effective "metallic layer" width $d_1$ of PMMA stripes was varied, while the width of effective "dielectric layers" (the gold/vacuum portions of the interface) $d_2$ was kept constant. Since eqs.(14,15) are approximations only, in the experiments presented in Fig.4 we have used a PMMA bi-grating geometry and varied its periodicity along both $x$ and $y$ directions within a broad range of parameters in order to achieve the presumed "end of time" conditions. Therefore, we followed the "combinatorial approach to metamaterials discovery" as described in ref.[33]. The bi-grating structures were defined using a Raith E-line electron beam lithography (EBL) system with ~70 nm spatial resolution. The written structures were subsequently developed using a 3:1 IPA/MIBK solution (Microchem) as developer. The fabricated structures were studied using an optical microscope under illumination with 488 nm

Argon ion laser, as described in [22]. It appears that the optical images of field distribution over the sample surface do indicate considerable field enhancement near the presumed plasmonic "end of time" events, as indicated by an arrow in Fig.4(f).

In conclusion, we have examined metamaterial optics at the boundaries of hyperbolic metamaterials. If the boundary is perpendicular to the space-like direction in the metamaterial, an effective Rindler horizon may be observed, which produces Hawking radiation. On the other hand, if the boundary is perpendicular to the time-like direction an unusual physics situation is created, which can be called "the end of time". It appears that in the lossless approximation electromagnetic field diverges at the interface in both situations. Experimental observations of the "end of time" using plasmonic metamaterials confirm this conclusion.

**Figure Captions**

**Figure 1.** Schematic views of “wired” (a) and “layered” (b) hyperbolic metamaterials. (c) Hyperbolic dispersion relation is illustrated as a surface of constant frequency in k-space. When $z$ coordinate is “timelike”, $k_z$ behaves as effective “energy”.

**Figure 2.** (a) Schematic representation of the “end of time” model in metamaterials: the spatial coordinate $z$ does not behave as a “timelike” variable any more, and the continuous “flow of time” is suddenly interrupted at the interface of two metamaterials. (b) Metric signature change across a “spacelike” direction leads to appearance of a Rindler horizon.

**Figure 3.** Numerical simulations of the “end of time” transition via $\varepsilon_2$=0. The field plot (a) and its cross section (b) illustrate field divergence at the interface.

**Figure 4.** Experimental observation of the “end of time event” in a plasmonic hyperbolic metamaterial illuminated with 488 nm light.

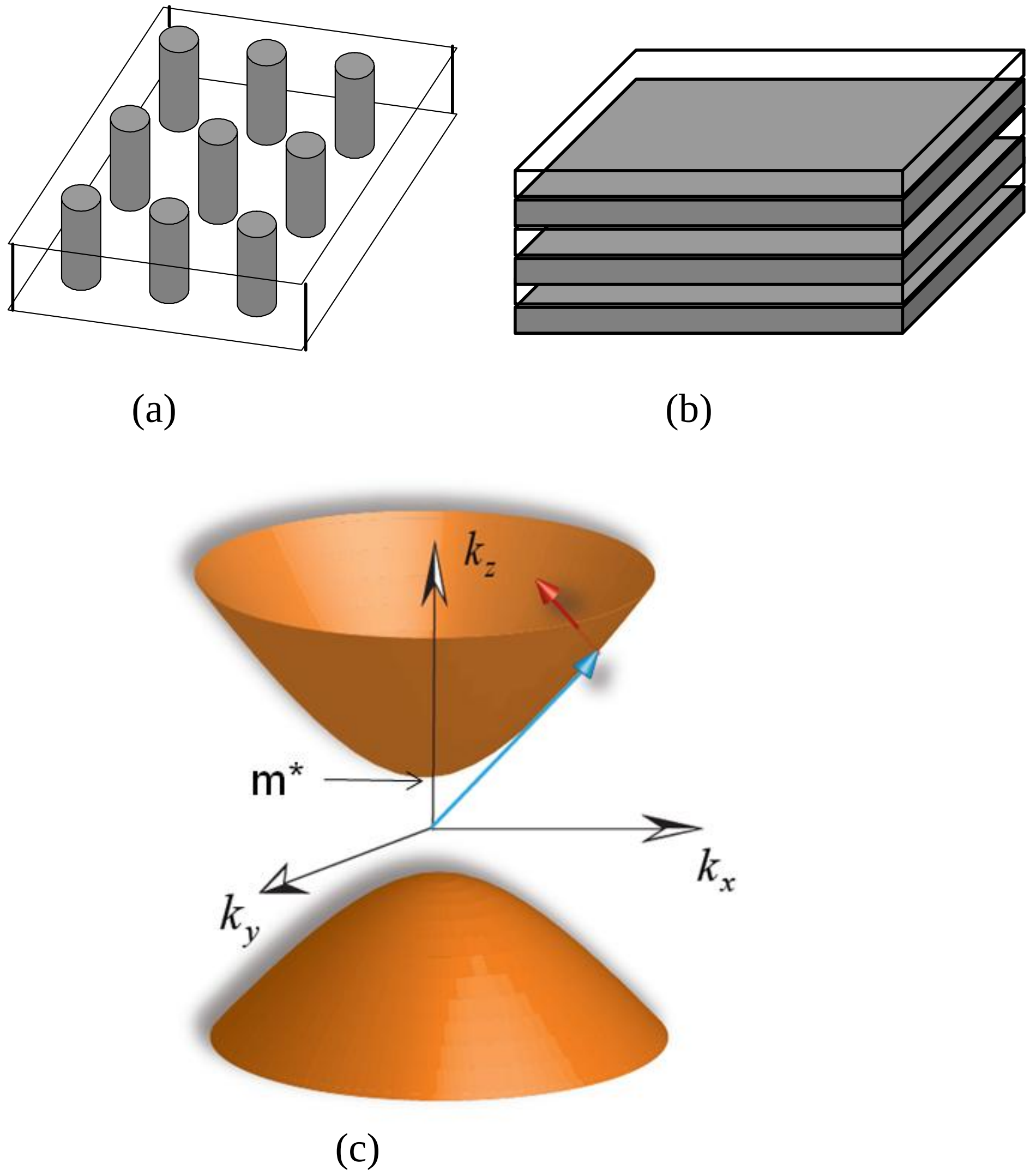

Fig.1

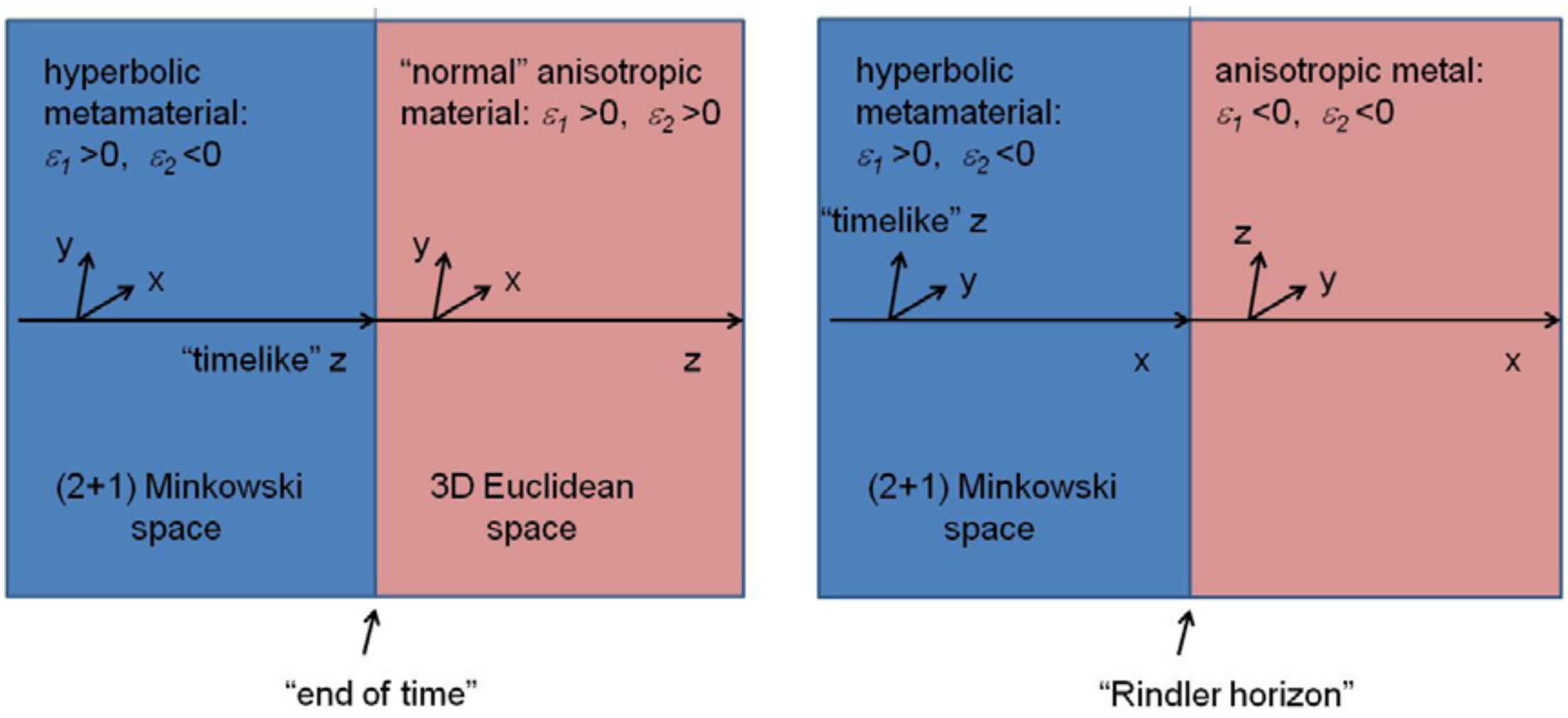

hyperbolic metamaterial:
$\varepsilon_1$ >0, $\varepsilon_2$ <0
y
x
"timelike" z
(2+1) Minkowski space
"normal" anisotropic material: $\varepsilon_1$ >0, $\varepsilon_2$ >0
y
x
z
3D Euclidean space
"end of time"
hyperbolic metamaterial:
$\varepsilon_1$ >0, $\varepsilon_2$ <0
"timelike" z
y
x
(2+1) Minkowski space
anisotropic metal:
$\varepsilon_1$ <0, $\varepsilon_2$ <0
z
y
x
"Rindler horizon"

(a) (b)

Fig.2

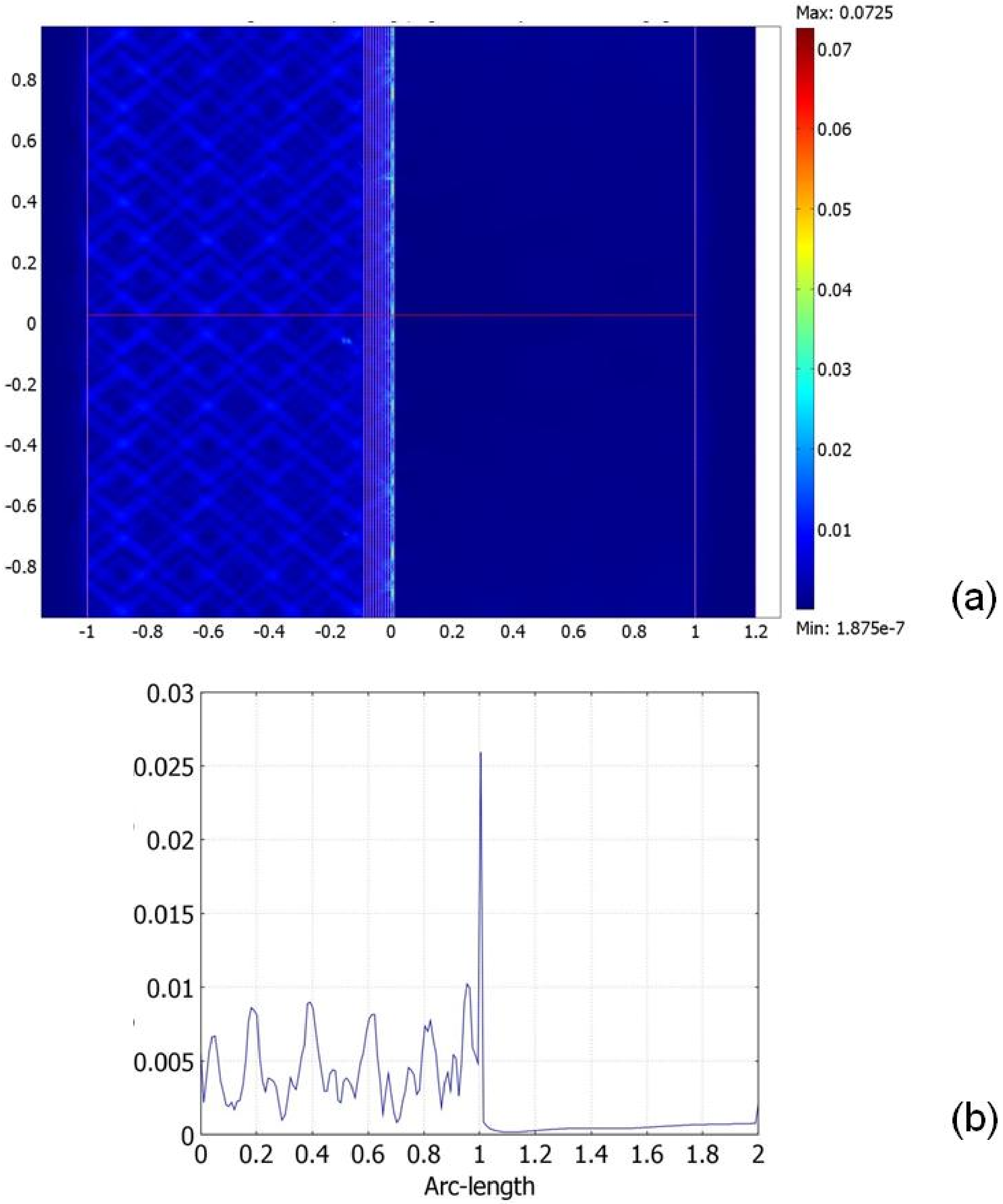

Fig.3

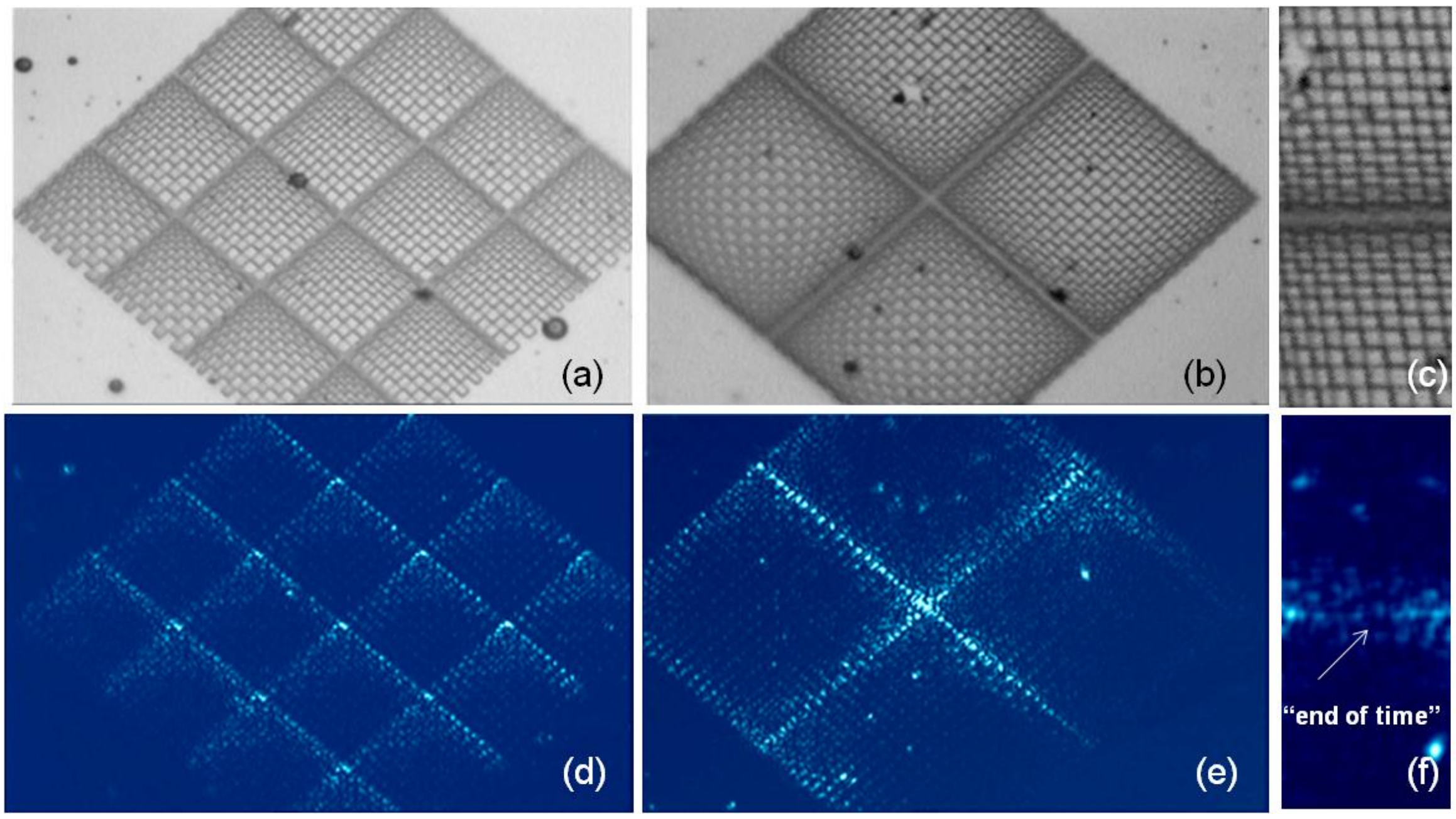

Fig.4